\documentclass[12pt]{article}
\usepackage{amsmath,amssymb}
\newcommand{\eqdef}{\stackrel{\text{def}}{=}}
\newcommand{\n}{\nonumber \\}
\newcommand{\bm}{\boldsymbol}
\newcommand{\ignore}[1]{}

\allowdisplaybreaks[4]

\begin{document}

\baselineskip=20pt

\newfont{\elevenmib}{cmmib10 scaled\magstep1}
\newcommand{\preprint}{
   \begin{flushleft}
     \elevenmib Yukawa\, Institute\, Kyoto\\
   \end{flushleft}\vspace{-1.3cm}
   \begin{flushright}\normalsize \sf
     DPSU-09-5\\
     YITP-09-69\\
   \end{flushright}}
\newcommand{\Title}[1]{{\baselineskip=26pt
   \begin{center} \Large \bf #1 \\ \ \\ \end{center}}}
\newcommand{\Author}{\begin{center}
   \large \bf S.~Odake${}^a$ and R.~Sasaki${}^b$ \end{center}}
\newcommand{\Address}{\begin{center}
     $^a$ Department of Physics, Shinshu University,\\
     Matsumoto 390-8621, Japan\\
     ${}^b$ Yukawa Institute for Theoretical Physics,\\
     Kyoto University, Kyoto 606-8502, Japan
   \end{center}}
\newcommand{\Accepted}[1]{\begin{center}
   {\large \sf #1}\\ \vspace{1mm}{\small \sf Accepted for Publication}
   \end{center}}

\preprint
\thispagestyle{empty}
\bigskip\bigskip\bigskip

\Title{Infinitely many shape invariant potentials and
cubic identities of the Laguerre and Jacobi polynomials}
\Author

\Address

\begin{abstract}
We provide analytic proofs for the shape invariance of the recently
discovered (Odake and Sasaki, Phys.\,\,Lett. {\bf B679} (2009) 414-417)
two families of infinitely many exactly solvable one-dimensional
quantum mechanical potentials.
These potentials are obtained by deforming the well-known radial
oscillator potential or the Darboux-P\"oschl-Teller potential
by a degree $\ell$ ($\ell=1,2,\ldots$) eigenpolynomial.
The shape invariance conditions are attributed to new polynomial
identities of degree $3\ell$ involving cubic products of the Laguerre
or Jacobi polynomials. These identities are proved elementarily by
combining simple identities.
\end{abstract}

\section{Introduction}
\setcounter{equation}{0}

In a previous Letter \cite{os16}, two sets of infinitely many shape
invariant \cite{genden} potentials were derived by deforming the
radial oscillator potential \cite{infhul,susyqm} and the
Darboux-P\"oschl-Teller (DPT) potential \cite{darboux,PT} in terms
of a polynomial eigenfunction of degree $\ell$ ($\ell=1,2,\ldots$).
As the main part of the eigenfunctions of these exactly solvable
quantum mechanical systems, the exceptional ($X_{\ell}$) Laguerre
and Jacobi polynomials were obtained \cite{os16}.
The lowest ($\ell=1$) examples, the $X_1$ Laguerre and Jacobi polynomials,
are equivalent to those introduced in the pioneering work of Gomez-Ullate
et al \cite{gomez, gomez2} within the Sturm-Liouville theory.
The reformulation in the framework of quantum mechanics and
shape-invariant potentials was done by Quesne et al \cite{quesne,BQR}.
By construction these new orthogonal polynomials satisfy a second order
differential equation (the Schr\"odinger equation) without contradicting
Bochner's theorem \cite{bochner}, since they start at degree $\ell$,
($\ell=1,2,\ldots$) instead of the degree zero constant term.

Here we present analytical proofs of the main assertion that these
deformed potentials are indeed shape invariant, which could not be
given in the Letter \cite{os16} due to the lack of space.
For both the deformed radial oscillator potential and deformed
trigonometric/hyperbolic DPT potentials, the condition for shape
invariance for each $\ell$ is satisfied if a certain identity involving
cubic products of the Laguerre or Jacobi polynomials holds.
To the best of our knowledge, these infinitely many identities were
not presented before. We show that these identities are derived by
combining several elementary relations among the Laguerre or Jacobi
polynomials.

This paper is organised as follows.
In section two the general setting of shape invariance and the
deformation of a potential in terms of a polynomial eigenfunction is
recapitulated together with many preparatory materials.
In section three we show that the condition for shape invariance
for each $\ell$ is attributed to a certain polynomial identity of
degree $3\ell$ involving cubic products of the Laguerre or Jacobi
polynomials of various parameters, \eqref{lagidenfin} and
\eqref{jacidenfin1}.
Then we show that these identities are simply derived by combining
simple relations among the Laguerre polynomials
\eqref{lemma1}--\eqref{lemma2} and Jacobi polynomials
\eqref{lemma3}--\eqref{lemma4}.
The final section is for a summary and comments on related results
to be published in a forthcoming paper.

\section{General setting}
\setcounter{equation}{0}

As in \cite{os16}, let us start with a generic one-dimensional
quantum mechanical system in a factorised form:
\begin{align}
  &\mathcal{H}=\mathcal{A}^{\dagger}\mathcal{A}=p^2+U(x),\quad
  p=-i\partial_x,
  \qquad U(x)\eqdef\bigl(\partial_xw(x)\bigr)^2+\partial_x^2w(x),\\
  &\mathcal{A}\eqdef\partial_x-\partial_xw(x),\quad
  \mathcal{A}^{\dagger}=-\partial_x-\partial_xw(x).
\end{align}
Here we call a real and smooth function $w(x)$ a {\em prepotential\/}.
It parametrises the groundstate wavefunction $\phi_0(x)$,
which has {\em no node\/} and can be chosen real and positive,
$\phi_0(x)=e^{w(x)}$. It is trivial to verify $\mathcal{A}\phi_0(x)=0$
and $\mathcal{H}\phi_0(x)=0$.

{\em Shape invariance\/} is realised by specific dependence of
the potential, or the prepotential on a set of parameters
$\bm{\lambda}=(\lambda_1,\lambda_2,\ldots)$, to be denoted by
$w(x;\bm{\lambda})$, $\mathcal{A}(\bm{\lambda})$,
$\mathcal{H}(\bm{\lambda})$, $\mathcal{E}_n(\bm{\lambda})$, etc.
The shape invariance condition to be discussed in this paper is
written simply:
\begin{align}
  \mathcal{A}(\bm{\lambda})\mathcal{A}(\bm{\lambda})^{\dagger}
  &=\mathcal{A}(\bm{\lambda}+\bm{\delta})^{\dagger}
  \mathcal{A}(\bm{\lambda}+\bm{\delta})
  +\mathcal{E}_1(\bm{\lambda}),
  \label{shapeinv1}\\
  \text{or}\quad \bigl(\partial_xw(x;\bm{\lambda})\bigr)^2
  -\partial_x^2w(x;\bm{\lambda})
  &=\bigl(\partial_xw(x;\bm{\lambda}+\bm{\delta})\bigr)^2
  +\partial_x^2w(x;\bm{\lambda}+\bm{\delta})+\mathcal{E}_1(\bm{\lambda}),
  \label{shapeinv2}
\end{align}
in which $\bm{\delta}$ is a certain shift of the parameters.
Then the entire set of discrete eigenvalues and the corresponding
eigenfunctions of $\mathcal{H}=\mathcal{H}(\bm{\lambda})$,
\begin{align}
  \mathcal{H}(\bm{\lambda})\phi_n(x;\bm{\lambda})
  &=\mathcal{E}_n(\bm{\lambda})\phi_n(x;\bm{\lambda}),\qquad\qquad
  n=0,1,2,\ldots,\\
  \phi_n(x;\bm{\lambda})&=\phi_0(x;\bm{\lambda})P_n(\eta(x);\bm{\lambda}),
  \label{eigenfun}
\end{align}
is determined algebraically \cite{genden,susyqm,crum,os4}:
\begin{align}
  \mathcal{E}_n(\bm{\lambda})
  &=\sum_{k=0}^{n-1}\mathcal{E}_1(\bm{\lambda}+k\bm{\delta}),
  \label{genformula0}\\
  \phi_n(x;\bm{\lambda})&\propto
  \mathcal{A}(\bm{\lambda})^{\dagger}
  \mathcal{A}(\bm{\lambda}+\bm{\delta})^{\dagger}\cdots
  \mathcal{A}(\bm{\lambda}+(n-1)\bm{\delta})^{\dagger}
  \times e^{w(x;\bm{\lambda}+n\bm{\delta})}.
  \label{genformula}
\end{align}
The polynomial eigenfunction $P_n(\eta(x);\bm{\lambda})$,
which is the Laguerre or Jacobi polynomial in $\eta(x)$, satisfies
\begin{equation}
  -\partial_x^2P_n(\eta(x);\bm{\lambda})
  -2\partial_xw(x;\bm{\lambda})\partial_xP_n(\eta(x);\bm{\lambda})
  =\mathcal{E}_n(\bm{\lambda})P_n(\eta(x);\bm{\lambda}).
  \label{Pndiffeq}
\end{equation}
Here $\eta(x)$ is a function of $x$ called the
{\em sinusoidal coordinate\/} \cite{os7}.

In \cite{os16} a shape invariant prepotential
$w(x;\bm{\lambda})=w_0(x;\bm{\lambda})$
is {\em deformed\/} by a polynomial eigenfunction $\xi_{\ell}$ of its
Hamiltonian to produce another shape invariant prepotential $w_{\ell}$
($\ell=1,2,\ldots.$):
\begin{equation}
  w_{\ell}(x;\bm{\lambda})\eqdef w_0(x;\bm{\lambda}+\ell\bm{\delta})
  +\log\frac{\xi_{\ell}(\eta(x);\bm{\lambda}+\bm{\delta})}
  {\xi_{\ell}(\eta(x);\bm{\lambda})},
  \label{wl}
\end{equation}
in which $\xi_{\ell}$ is related to the polynomial eigenfunction $P_n$
above \eqref{eigenfun}.
It should be noted that the normalisation of the polynomial
$\xi_{\ell}(x;\bm{\lambda})$ is irrelevant to the deformation.
The $\ell=0$ case corresponds to the original system.
The $\ell$-th Hamiltonian and eigenfunctions, etc are given by
\begin{align}
  &\mathcal{A}_{\ell}(\bm{\lambda})\eqdef\partial_x
  -\partial_xw_{\ell}(x;\bm{\lambda}),\quad
  \mathcal{A}_{\ell}(\bm{\lambda})^{\dagger}
  =-\partial_x-\partial_xw_{\ell}(x;\bm{\lambda}),
  \label{Aldef}\\
  &\mathcal{H}_{\ell}(\bm{\lambda})\eqdef
  \mathcal{A}_{\ell}(\bm{\lambda})^{\dagger}\mathcal{A}_{\ell}(\bm{\lambda}),
  \label{Hldef}\\
  &\mathcal{H}_{\ell}(\bm{\lambda})\phi_{\ell,n}(x;\bm{\lambda})
  =\mathcal{E}_n(\bm{\lambda}+\ell\bm{\delta})\phi_{\ell,n}(x;\bm{\lambda}),\\
  &\phi_{\ell,n}(x;\bm{\lambda})
  =\psi_{\ell}(x;\bm{\lambda})P_{\ell,n}(\eta(x);\bm{\lambda}),\quad
  \psi_{\ell}(x;\bm{\lambda})\eqdef
  \frac{e^{w_0(x;\bm{\lambda}+\ell\bm{\delta})}}
  {\xi_{\ell}(\eta(x);\bm{\lambda})}.
  \label{genmeasure}
\end{align}
The orthogonality of the eigenfunctions of the Hamiltonian
$\mathcal{H}_{\ell}(\bm{\lambda})$ reads
\begin{equation}
  \int\psi_{\ell}(x;\bm{\lambda})^2P_{\ell,n}(\eta(x);\bm{\lambda})
  P_{\ell,m}(\eta(x);\bm{\lambda})dx
  =h_{\ell,n}(\bm{\lambda})\delta_{n\,m},
  \quad h_{\ell,n}(\bm{\lambda})>0,
\end{equation}
in which $\psi_{\ell}(x;\bm{\lambda})^2$ is the orthogonality measure and
$P_{\ell,n}(x;\bm{\lambda})$ is the $n$-th member of the {\em exceptional}
($X_{\ell}$) {\em orthogonal polynomial}.  It is expressed in terms of
$\xi_{\ell}(x)$'s and $P_n(x)$'s as shown in \eqref{PlnLag},
\eqref{Plnjac} and \eqref{hypPlnjac}.

In this paper we will demonstrate that the deformed prepotential $w_{\ell}$
($\ell=1,2,\ldots$) actually satisfies the shape invariance condition
\begin{equation}
  \Delta_{\ell}(x;\bm{\lambda})=0,
  \label{Deltazero}
\end{equation}
in which $\Delta_{\ell}(x;\bm{\lambda})$ is defined by
\begin{equation}
  \Delta_{\ell}(x;\bm{\lambda})\eqdef
  \bigl(\partial_x w_{\ell}(x;\bm{\lambda})\bigr)^2
  -\partial_x^2w_{\ell}(x;\bm{\lambda})
  -\bigl(\partial_xw_{\ell}(x;\bm{\lambda}+\bm{\delta})\bigr)^2
  -\partial_x^2w_{\ell}(x;\bm{\lambda}+\bm{\delta})
  -\mathcal{E}_1(\bm{\lambda}+\ell\bm{\delta}),
  \label{Deltadef}
\end{equation}
for the three cases, the radial oscillator and the trigonometric/hyperbolic
DPT discussed in \cite{os16}.
The proof consists of two steps.
Firstly in section three we transform, by utilising the differential
equations for $\xi_{\ell}(\eta(x);\bm{\lambda})$ \eqref{xildiffeq} etc,
the shape invariance condition \eqref{Deltadef} into an identity
involving products of three Laguerre or Jacobi polynomials of various
parameters, \eqref{lagidenfin} and \eqref{jacidenfin1}.
The trigonometric and hyperbolic DPT lead to the same identities.
Secondly these cubic identities are proved by combining simple identities
among the Laguerre or Jacobi polynomials of neighbouring degrees $n$,
$n-1$ and parameters $\alpha$, $\alpha\pm1$, $\beta$, $\beta\pm1$.

Here we show various data necessary for the proof.
They are recapitulated from \cite{os16}.
\begin{enumerate}
\item[1] {\bf radial oscillator}:
\begin{align}
  &\bm{\lambda}\eqdef g,\quad\bm{\delta}=1,\quad g>0,
  \label{raddat1}\\
  &\mathcal{E}_n(\bm{\lambda})=4n,\quad\eta(x)\eqdef x^2,\quad 0<x<\infty,
  \label{ratdomain}\\
  &\phi_0(x;\bm{\lambda})\eqdef e^{-\tfrac{x^2}{2}}x^g\Leftrightarrow
   w_0(x;\bm{\lambda})\eqdef-\frac{x^2}{2}+g\log x,\\
  &P_n(x;\bm{\lambda})\eqdef L_n^{(g-\tfrac{1}{2})}(x),\quad
  \xi_{\ell}(x;\bm{\lambda})\eqdef L_{\ell}^{(g+\ell-\tfrac{3}{2})}(-x),\\
  &P_{\ell,n}(x;\bm{\lambda})=\xi_{\ell}(x;g+1)P_n(x;g+\ell)
  -\xi_{\ell-1}(x;g+2)P_{n-1}(x;g+\ell).
  \label{PlnLag}
\end{align}
\item[2] {\bf trigonometric DPT}:
\begin{align}
  &\bm{\lambda}\eqdef(g,h),\quad\bm{\delta}=(1,1),\quad h>g>0,
  \label{trigdat1}\\
  &\mathcal{E}_n(\bm{\lambda})=4n(n+g+h),\quad
  \eta(x)\eqdef\cos2x,\quad 0<x<\frac{\pi}{2},
  \label{trigdomain}\\
  &\phi_0(x;\bm{\lambda})\eqdef(\sin x)^g(\cos x)^h\Leftrightarrow
  w_0(x;\bm{\lambda})=g\log\sin x+h\log\cos x,\\
  &P_n(x;\bm{\lambda})\eqdef P_n^{(g-\frac12,\,h-\frac12)}(x),\quad
  \xi_{\ell}(x;\bm{\lambda})\eqdef
  P_{\ell}^{(-g-\ell-\frac12,\,h+\ell-\frac32)}(x),\\
  &P_{\ell,n}(x;\bm{\lambda})\eqdef
  a_{\ell,n}(x;\bm{\lambda})P_n(x;\bm{\lambda}+\ell\bm{\delta})
  +b_{\ell,n}(x;\bm{\lambda})P_{n-1}(x;\bm{\lambda}+\ell\bm{\delta}),
  \label{Plnjac}\\
  &a_{\ell,n}(x;\bm{\lambda})\eqdef\xi_{\ell}(x;g+1,h+1)
  +\frac{2n(-g+h+\ell-1)\,\xi_{\ell-1}(x;g,h+2)}
  {(-g+h+2\ell-2)(g+h+2n+2\ell-1)}\n[2pt]
  &\phantom{a_{\ell,n}(x;\bm{\lambda})\eqdef\xi_{\ell}(x;g+1,h+1)}
  -\frac{n(2h+4\ell-3)\,\xi_{\ell-2}(x;g+1,h+3)}
  {(2g+2n+1)(-g+h+2\ell-2)},\\
  &b_{\ell,n}(x;\bm{\lambda})\eqdef
  \frac{(-g+h+\ell-1)(2g+2n+2\ell-1)}{(2g+2n+1)(g+h+2n+2\ell-1)}\,
  \xi_{\ell-1}(x;g,h+2).
  \label{PlnJac}
\end{align}
\item[3] {\bf hyperbolic DPT}:
\begin{align}
  &\bm{\lambda}\eqdef(g,h),\quad\bm{\delta}=(1,-1),\quad h>g>0,\quad
  \ell<n_B\eqdef[\tfrac12(h-g)]',
  \label{hypdat1}\\
  &\mathcal{E}_n(\bm{\lambda})=4n(h-g-n),\quad
  \eta(x)=\cosh2x,\quad 0<x<\infty,
  \label{hypdomain}\\
  &\phi_0(x;\bm{\lambda})\eqdef(\sinh x)^g(\cosh x)^{-h}\Leftrightarrow
  w_0(x;\bm{\lambda})=g\log\sinh x-h\log\cosh x,\\
  &P_n(x;\bm{\lambda})\eqdef P_n^{(g-\frac12,\,-h-\frac12)}(x),\quad
  \xi_{\ell}(x;\bm{\lambda})\eqdef
  P_{\ell}^{(-g-\ell-\frac12,\,-h+\ell-\frac32)}(x),\\
  &P_{\ell,n}(x;\bm{\lambda})\eqdef
  a_{\ell,n}(x;\bm{\lambda})P_n(x;\bm{\lambda}+\ell\bm{\delta})
  +b_{\ell,n}(x;\bm{\lambda})P_{n-1}(x;\bm{\lambda}+\ell\bm{\delta}),
  \label{hypPlnjac}\\
  &a_{\ell,n}(x;\bm{\lambda})\eqdef\xi_{\ell}(x;g+1,h-1)
  +\frac{2n(-g-h+\ell-1)\,\xi_{\ell-1}(x;g,h-2)}
  {(-g-h+2\ell-2)(g-h+2n+2\ell-1)}\n[2pt]
  &\phantom{a_{\ell,n}(x;\bm{\lambda})\eqdef\xi_{\ell}(x;g+1,h-1)}
  -\frac{n(-2h+4\ell-3)\,\xi_{\ell-2}(x;g+1,h-3)}
  {(2g+2n+1)(-g-h+2\ell-2)},\\
  &b_{\ell,n}(x;\bm{\lambda})\eqdef
  \frac{(-g-h+\ell-1)(2g+2n+2\ell-1)}{(2g+2n+1)(g-h+2n+2\ell-1)}\,
  \xi_{\ell-1}(x;g,h-2).
  \label{PlnJac2}
\end{align}
\end{enumerate}
In \eqref{hypdat1} $[x]'$ denotes the greatest integer not equal or
exceeding $x$.
Here $L^{(\alpha)}_n(x)$ is the Laguerre polynomial and
$P_n^{(\alpha,\,\beta)}(x)$ is the Jacobi polynomial.
The polynomial eigenfunction $P_{\ell,n}(x;\bm{\lambda})$ \eqref{PlnLag},
\eqref{Plnjac} or \eqref{hypPlnjac} is a degree $\ell+n$ polynomial in $x$.
In a future publication \cite{hos} we will present equivalent forms of the
$X_{\ell}$ polynomials $P_{\ell,n}(x;\bm{\lambda})$ which appear much simpler
than those given in \eqref{PlnLag}, \eqref{Plnjac} and \eqref{hypPlnjac}.
Needless to say that the deforming polynomials
$\xi_{\ell}(\eta(x);\bm{\lambda})$ and
$\xi_{\ell}(\eta(x);\bm{\lambda}+\bm{\delta})$ are of the same sign
in the domain for all the three cases. In other words, the deforming
polynomials do not have a zero in the domains listed in \eqref{ratdomain},
\eqref{trigdomain} and \eqref{hypdomain}, respectively.
To see this we use the expansion formula of the Laguerre and Jacobi
polynomials:
\begin{align}
  L_n^{(\alpha)}(x)&=\frac{1}{n!}
  \sum_{k=0}^n\frac{(-n)_k}{k!}(\alpha+k+1)_{n-k}x^k,
  \label{Lagexp}\\
  P_n^{(\alpha,\,\beta)}(x)&=\frac{(\alpha+1)_n}{n!}
  \sum_{k=0}^n\frac{1}{k!}\frac{(-n)_k(n+\alpha+\beta+1)_k}{(\alpha+1)_k}
  \Bigl(\frac{1-x}{2}\Bigr)^k.
  \label{Jacexp}
\end{align}
For the radial oscillator we obtain
\begin{equation}
  \xi_{\ell}(\eta(x);\bm{\lambda})
  =\sum_{k=0}^{\ell}\frac{(g+\ell+k-\frac12)_{\ell-k}}{k!\,(\ell-k)!}
  \,x^{2k}>0,
\end{equation}
and for the trigonometric/hyperbolic DPT
\begin{equation}
  (-1)^{\ell}\xi_{\ell}(\eta(x);\bm{\lambda})=\left\{
  \begin{array}{l}
  {\displaystyle\frac{(g+\frac12)_{\ell}}{\ell!}\sum_{k=0}^{\ell}
  \frac{(\ell-k+1)_k(h-g+\ell-1)_k}{k!\,(g+\ell-k+\frac12)_k}
  (\sin x)^{2k}>0},\\
  {\displaystyle\frac{(g+\frac12)_{\ell}}{\ell!}\sum_{k=0}^{\ell}
  \frac{(\ell-k+1)_k(g+h+2-\ell-k)_k}{k!\,(g+\ell-k+\frac12)_k}
  (\sinh x)^{2k}>0},
  \end{array}\right.
\end{equation}
because each term in summation is positive.
This guarantees the positive definiteness of the orthogonality measure
$\psi_{\ell}(x;\bm{\lambda})^2$ \eqref{genmeasure} and the singularity
free structure as well as the hermiticity (self-adjointness) of the
Hamiltonian.
The oscillation theorem for the one-dimensional quantum mechanical
systems dictates that the $n$-th excited state polynomial eigenfunction
$P_{\ell,n}(\eta(x);\bm{\lambda})$ has $n$ zeros in the domain
\eqref{ratdomain}, \eqref{trigdomain} or \eqref{hypdomain}, although
it is a degree $\ell+n$ polynomial in $\eta$.

\bigskip
Let us note that the terms of the form of the complete square
$\bigl(\partial_x\xi_{\ell}(\eta(x);\bm{\lambda})/
\xi_{\ell}(\eta(x);\bm{\lambda})\bigr)^2$,
$\bigl(\partial_x\xi_{\ell}(\eta(x);\bm{\lambda}+\bm{\delta})/
\xi_{\ell}(\eta(x);\bm{\lambda}+\bm{\delta})\bigr)^2$ and
$\bigl(\partial_x\xi_{\ell}(\eta(x);\bm{\lambda}+2\bm{\delta})/
\xi_{\ell}(\eta(x);\bm{\lambda}+2\bm{\delta})\bigr)^2$,
cancel out in $\Delta_{\ell}(x;\bm{\lambda})$ \eqref{Deltadef}.
Then we use the fact that, corresponding to \eqref{Pndiffeq},
the deforming polynomial $\xi_{\ell}(\eta(x);\bm{\lambda})$ also
satisfies a second order linear differential equation
\begin{equation}
  -\partial_x^2\xi_{\ell}(\eta(x);\bm{\lambda})
  -2\partial_x\widetilde{w}_0(x;\bm{\lambda},\ell)
  \partial_x\xi_{\ell}(\eta(x);\bm{\lambda})
  =\widetilde{\mathcal{E}}_{\ell}(\bm{\lambda})
  \xi_{\ell}(\eta(x);\bm{\lambda}),
  \label{xildiffeq}
\end{equation}
where $\widetilde{w}_0(x;\bm{\lambda},\ell)$ and
$\widetilde{\mathcal{E}}_{\ell}(\bm{\lambda})$ are
\begin{align}
  \widetilde{w}_0(x;\bm{\lambda},\ell)&=\left\{
  \begin{array}{ll}
  \tfrac12x^2+(g+\ell-1)\log x&\,:\text{radial osci.}\\
  w_0(x;-g-\ell,h+\ell-1)&\,:\text{trig.DPT}\\
  w_0(x;-g-\ell,h-\ell+1)&\,:\text{hyper.DPT},
  \end{array}\right.\\
  \widetilde{\mathcal{E}}_{\ell}(\bm{\lambda})&=\left\{
  \begin{array}{ll}
  -4\ell&\hspace*{6mm}:\text{radial osci.}\\
  \mathcal{E}_{\ell}(-g-\ell,h+\ell-1)&\hspace*{6mm}:\text{trig.DPT}\\
  \mathcal{E}_{\ell}(-g-\ell,h-\ell+1)&\hspace*{6mm}:\text{hyper.DPT}.
  \end{array}\right.
\end{align}
Substituting \eqref{wl} into \eqref{Deltadef} and using the shape
invariance of the undeformed system
$\Delta_0(x;\bm{\lambda}+\ell\bm{\delta})=0$ and \eqref{xildiffeq},
we obtain
\begin{align}
  &\Delta_{\ell}(x;\bm{\lambda})\times
  \frac12\,\xi_{\ell}(\eta;\bm{\lambda})
  \,\xi_{\ell}(\eta;\bm{\lambda}+\bm{\delta})
  \,\xi_{\ell}(\eta;\bm{\lambda}+2\bm{\delta})\n
  =\,&\partial_x\widetilde{w}_0(x;\bm{\lambda}+2\bm{\delta},\ell)
  \,\partial_x\xi_{\ell}(\eta;\bm{\lambda}+2\bm{\delta})
  \,\xi_{\ell}(\eta;\bm{\lambda})
  \,\xi_{\ell}(\eta;\bm{\lambda}+\bm{\delta})\n
  &-\partial_x\widetilde{w}_0(x;\bm{\lambda},\ell)
  \,\partial_x\xi_{\ell}(\eta;\bm{\lambda})
  \,\xi_{\ell}(\eta;\bm{\lambda}+\bm{\delta})
  \,\xi_{\ell}(\eta;\bm{\lambda}+2\bm{\delta})\n
  &+\tfrac12\bigl(\widetilde{\mathcal{E}}_{\ell}(\bm{\lambda}+2\bm{\delta})
  -\widetilde{\mathcal{E}}_{\ell}(\bm{\lambda})\bigr)
  \,\xi_{\ell}(\eta;\bm{\lambda})
  \,\xi_{\ell}(\eta;\bm{\lambda+\delta})
  \,\xi_{\ell}(\eta;\bm{\lambda}+2\bm{\delta})\n
  &+\partial_xw_0(x;\bm{\lambda}+\ell\bm{\delta})
  \bigl(\partial_x\xi_{\ell}(\eta;\bm{\lambda}+\bm{\delta})
  \,\xi_{\ell}(\eta;\bm{\lambda})
  -\partial_x\xi_{\ell}(\eta;\bm{\lambda})
  \,\xi_{\ell}(\eta;\bm{\lambda}+\bm{\delta})\bigr)
  \xi_{\ell}(\eta;\bm{\lambda}+2\bm{\delta})\n
  &-\partial_xw_0(x;\bm{\lambda}+\ell\bm{\delta}+\bm{\delta})
  \bigl(\partial_x\xi_{\ell}(\eta;\bm{\lambda}+2\bm{\delta})
  \xi_{\ell}(\eta;\bm{\lambda}+\bm{\delta})
  -\partial_x\xi_{\ell}(\eta;\bm{\lambda}+\bm{\delta})
  \xi_{\ell}(\eta;\bm{\lambda}+2\bm{\delta})\bigr)
  \xi_{\ell}(\eta;\bm{\lambda})\n
  &-\partial_x\xi_{\ell}(\eta;\bm{\lambda})
  \,\partial_x\xi_{\ell}(\eta;\bm{\lambda}+\bm{\delta})
  \,\xi_{\ell}(\eta;\bm{\lambda}+2\bm{\delta})
  +\partial_x\xi_{\ell}(\eta;\bm{\lambda}+\bm{\delta})
  \,\partial_x\xi_{\ell}(\eta;\bm{\lambda}+2\bm{\delta})
  \,\xi_{\ell}(\eta;\bm{\lambda}),
  \label{shapeD2}
\end{align}
where $\eta=\eta(x)$.
It is essential that the r.h.s. of \eqref{shapeD2} is now a polynomial
in $\eta(x)$, because $\partial_x\eta(x)\partial_xw_0$ and
$\partial_x\eta(x)\partial_x\widetilde{w}_0$ and $(\partial_x\eta(x))^2$
are expressed by $\eta(x)$,
\begin{align}
  \partial_x\eta(x)\partial_xw_0(x;\bm{\lambda})&=
  2\times\left\{
  \begin{array}{ll}
  g-\eta(x)&\hspace*{23mm}:\text{radial osci.}\\
  -\bigl(g-h+(g+h)\eta(x)\bigr)&\hspace*{23mm}:\text{trig.DPT}\\
  g+h+(g-h)\eta(x)&\hspace*{23mm}:\text{hyper.DPT},
  \end{array}\right.
  \label{dw0}\\[3pt]
  \partial_x\eta(x) \partial_x\widetilde{w}_0(x;\bm{\lambda},\ell)&=
  2\times\left\{
  \begin{array}{ll}
  g+\ell-1+\eta(x)&:\text{radial osci.}\\
  g+h+2\ell-1+(g-h+1)\eta(x)&:\text{trig.DPT}\\
  -\bigl(g-h+2\ell-1+(g+h+1)\eta(x)\bigr)&:\text{hyper.DPT},
  \end{array}\right.\\[3pt]
  \bigl(\partial_x\eta(x)\bigr)^2&=
  4\times\left\{
  \begin{array}{ll}
  \eta(x)&\hspace*{42mm}:\text{radial osci.}\\
  1-\eta(x)^2&\hspace*{42mm}:\text{trig.DPT}\\
  -\bigl(1-\eta(x)^2\bigr)&\hspace*{42mm}:\text{hyper.DPT}.
  \end{array}\right.
\end{align}

\section{Cubic identities}
\setcounter{equation}{0}

The conditions for the shape invariance \eqref{Deltazero} are shown
to be equivalent to cubic identities involving the Laguerre or Jacobi
polynomials.

\subsection{radial oscillator}

We fix $\ell$ and use a new parameter $\alpha$ instead of $g$,
\begin{equation}
  \alpha\eqdef g+\ell-\tfrac12.
\end{equation}
By using the forward shift relation for the Laguerre polynomial,
\begin{equation}
  \partial_xL_n^{(\alpha)}(x)=-L_{n-1}^{(\alpha+1)}(x),
\end{equation}
the polynomial $\xi_{\ell}$ and its derivative are expressed as
\begin{align}
  \xi_{\ell}(\eta;\bm{\lambda})&=L_{\ell}^{(\alpha-1)}(-\eta),\quad
  \partial_x\xi_{\ell}(\eta;\bm{\lambda})
  =\partial_x\eta\ L_{\ell-1}^{(\alpha)}(-\eta),
\end{align}
where $\eta=\eta(x)$.
After replacing $-\eta(x)$ with $x$ and dividing by 4, the condition
for the shape invariance \eqref{Deltazero} with \eqref{shapeD2}, is
transformed into the following polynomial identity of degree $3\ell$
in $x$ which contains products of three Laguerre polynomials of various
parameters:
\begin{align}
  0=&-x L_{\ell-1}^{(\alpha+2)}(x)
  L_{\ell}^{(\alpha-1)}(x)L_{\ell}^{(\alpha)}(x) 
  -\alpha L_{\ell-1}^{(\alpha)}(x)
  L_{\ell}^{(\alpha+1)}(x)L_{\ell}^{(\alpha)}(x)\n
  &+(x+\alpha+1)L_{\ell-1}^{(\alpha+1)}(x)
  L_{\ell}^{(\alpha+1)}(x)L_{\ell}^{(\alpha-1)}(x)\n
  &+x L_{\ell-1}^{(\alpha)}(x)
  L_{\ell-1}^{(\alpha+1)}(x)L_{\ell}^{(\alpha+1)}(x)
  -x L_{\ell-1}^{(\alpha+1)}(x)
  L_{\ell-1}^{(\alpha+2)}(x)L_{\ell}^{(\alpha-1)}(x).
  \label{lagidenfin}
\end{align}
To the best of our knowledge, this identity has not been reported before.
For $\ell=0$ this identity is trivial, since $L_{-1}^{(\alpha)}(x)=0$.
For lower $\ell$ it can be easily verified by direct calculation.

Below we will prove the identity \eqref{lagidenfin} for an arbitrary
positive integer $\ell$ by combining a few elementary relations among
the Laguerre polynomials of neighbouring degrees $n$ and $n-1$ and
neighbouring parameters $\alpha$, $\alpha\pm1$:
\begin{alignat}{2}
  &\text{Lemma (A)} \qquad
  &L_n^{(\alpha-1)}(x)+L_{n-1}^{(\alpha)}(x)&=L_n^{(\alpha)}(x),
  \label{lemma1}\\
  &\text{Lemma (B)} \qquad
  &x L_{n-1}^{(\alpha+1)}(x)-\alpha L_{n-1}^{(\alpha)}(x)
  &=-n L_n^{(\alpha-1)}(x),
  \label{lemma2}
\end{alignat}
which can be verified elementarily based on the expansion formula of
the Laguerre polynomial \eqref{Lagexp}.
The r.h.s. of the identity \eqref{lagidenfin} can be written as
\begin{align}
  \text{r.h.s. of \eqref{lagidenfin}}
  &=-x L_{\ell-1}^{(\alpha+2)}(x)L_{\ell}^{(\alpha-1)}(x)
  \bigl\{L_{\ell}^{(\alpha)}(x)+L_{\ell-1}^{(\alpha+1)}(x)\bigr\}\n
  &\quad +x L_{\ell-1}^{(\alpha+1)}(x)L_{\ell}^{(\alpha+1)}(x)
  \bigl\{L_{\ell}^{(\alpha-1)}(x)+L_{\ell-1}^{(\alpha)}(x)\bigr\}\n
  &\quad +L_{\ell}^{(\alpha+1)}(x)
  \bigl((\alpha+1)L_{\ell-1}^{(\alpha+1)}(x)L_{\ell}^{(\alpha-1)}(x)
  -\alpha L_{\ell-1}^{(\alpha)}(x)L_{\ell}^{(\alpha)}(x)\bigr)\n
  &=L_{\ell}^{(\alpha+1)}(x)\Bigl(L_{\ell}^{(\alpha)}(x)
  \bigl\{x L_{\ell-1}^{(\alpha+1)}(x)-\alpha L_{\ell-1}^{(\alpha)}(x)\bigr\}\n
  &\phantom{=L_{\ell}^{(\alpha+1)}(x)\Bigl(}
  +L_{\ell}^{(\alpha-1)}(x)
  \bigl\{-x L_{\ell-1}^{(\alpha+2)}(x)
  +(\alpha+1)L_{\ell-1}^{(\alpha+1)}(x)\bigr\}\Bigr)\n
  &=L_{\ell}^{(\alpha+1)}(x)\Bigl(
  L_{\ell}^{(\alpha)}(x)\bigl(-\ell L_{\ell}^{(\alpha-1)}(x)\bigr)
  +L_{\ell}^{(\alpha-1)}(x)\bigl(+\ell L_{\ell}^{(\alpha)}(x)\bigr)\Bigr)
  =0.
\end{align}
Here we have used Lemma (A) in the first two curly brackets $\{\cdots\}$
and Lemma (B) in the next two curly brackets.
This concludes the proof of the identity \eqref{lagidenfin}.

\subsection{trigonometric DPT}

We fix $\ell$ and use new parameters $\alpha$ and $\beta$ instead of
$g$ and $h$,
\begin{equation}
  \alpha\eqdef -g-\ell-\tfrac12,\quad \beta\eqdef h+\ell-\tfrac32.
  \label{trigDPTab}
\end{equation}
By using the forward shift relation for the Jacobi polynomial,
\begin{equation}
  \partial_xP_n^{(\alpha,\,\beta)}(x)
  =\tfrac12(n+\alpha+\beta+1)P_{n-1}^{(\alpha+1,\,\beta+1)}(x),
  \label{dPn=Pn-1}
\end{equation}
the polynomial $\xi_{\ell}$ and its derivative are expressed as
\begin{align}
  \xi_{\ell}(\eta;\bm{\lambda})
  &=P_{\ell}^{(\alpha,\,\beta)}(\eta),\quad
  \partial_x\xi_{\ell}(\eta;\bm{\lambda})
  =\partial_x\eta\ \tfrac12(\ell+\alpha+\beta+1)
  P_{\ell-1}^{(\alpha+1,\,\beta+1)}(\eta),
  \label{xil->P}
\end{align}
where $\eta=\eta(x)$.
The condition for the shape invariance \eqref{Deltazero} with
\eqref{shapeD2}, after replacing $\eta(x)$ with $x$ and dividing by
$-(\ell+\alpha+\beta+1)$, is simplified to a polynomial identity of
degree $3\ell$, which contains products of three Jacobi polynomials
of various parameters:
\begin{align}
  0=&\ 2(\alpha-1)(1+x)P_{\ell-1}^{(\alpha-1,\beta+3)}(x)
  P_{\ell}^{(\alpha,\beta)}(x)P_{\ell}^{(\alpha-1,\beta+1)}(x)\n
  &+2(\beta+1)(1-x)P_{\ell-1}^{(\alpha+1,\beta+1)}(x)
  P_{\ell}^{(\alpha-2,\beta+2)}(x)P_{\ell}^{(\alpha-1,\beta+1)}(x)\n
  &-2\bigl(\alpha(1+x)+(\beta+2)(1-x)\bigr)
  P_{\ell-1}^{(\alpha,\beta+2)}(x)
  P_{\ell}^{(\alpha,\beta)}(x)P_{\ell}^{(\alpha-2,\beta+2)}(x)\n
  &+(\ell+\alpha+\beta+1)(1-x^2)P_{\ell-1}^{(\alpha+1,\beta+1)}(x)
  P_{\ell-1}^{(\alpha,\beta+2)}(x)P_{\ell}^{(\alpha-2,\beta+2)}(x)\n
  &-(\ell+\alpha+\beta+1)(1-x^2)
  P_{\ell-1}^{(\alpha,\beta+2)}(x)P_{\ell-1}^{(\alpha-1,\beta+3)}(x)
  P_{\ell}^{(\alpha,\beta)}(x).
  \label{jacidenfin1}
\end{align}
To the best of our knowledge, this identity has not been reported before,
either. For $\ell=0$ this identity is trivial, since
$P_{-1}^{(\alpha,\beta)}(x)=0$. It is straightforward to verify this
identity for lower $\ell$  by direct calculation.

Below we will prove the identity \eqref{jacidenfin1} for an arbitrary
positive integer $\ell$ by combining a few elementary relations among
the Jacobi polynomials of neighbouring degrees $n$ and $n-1$ and
neighbouring parameters $\alpha\pm1,\beta\pm1$:
\begin{align}
  &\text{Lemma (C)}\n
  &\ \ 2(\alpha-1)P_n^{(\alpha-1,\beta)}(x)
  -(n+\alpha+\beta)(1-x)P_{n-1}^{(\alpha,\beta+1)}(x)
  =2(n+\alpha-1)P_n^{(\alpha-2,\beta+1)}(x),
  \label{lemma3}\\
  &\text{Lemma (D)}\n
  &\ 2(\beta+1) P_{n}^{(\alpha-1,\beta+1)}(x)
  \!+\!(n+\alpha+\beta+1)(1+x) P_{n-1}^{(\alpha,\beta+2)}(x)
  =2(n+\beta+1)P_n^{(\alpha,\beta)}(x).
  \label{lemma4}
\end{align}
It is straightforward to demonstrate Lemma (C) by using the expansion
formula for the Jacobi polynomials \eqref{Jacexp}.
By using the property
$P_n^{(\alpha,\,\beta)}(-x)=(-1)^nP_n^{(\beta,\,\alpha)}(x)$,
Lemma (D) is obtained from Lemma (C) with the replacements $x\to -x$,
$\alpha\to\beta+2$, $\beta\to\alpha-1$.
The r.h.s. of the identity \eqref{jacidenfin1} can be written as
\begin{align}
  \text{r.h.s. of \eqref{jacidenfin1}}
  &=(1+x)P_{\ell-1}^{(\alpha-1,\beta+3)}(x)P_{\ell}^{(\alpha,\beta)}(x)\n
  &\qquad\ \times\bigl\{2(\alpha-1)P_{\ell}^{(\alpha-1,\beta+1)}(x)
  -(\ell+\alpha+\beta+1)(1-x)P_{\ell-1}^{(\alpha,\beta+2)}(x)\bigr\}\n
  &\phantom{=}
  \ +(1-x)P_{\ell-1}^{(\alpha+1,\beta+1)}(x)P_{\ell}^{(\alpha-2,\beta+2)}(x)\n
  &\qquad\ \times\bigl\{2(\beta+1)P_{\ell}^{(\alpha-1,\beta+1)}(x)
  +(\ell+\alpha+\beta+1)(1+x)P_{\ell-1}^{(\alpha,\beta+2)}(x)\bigr\}\n
  &\phantom{=}
  \ -2\bigl(\alpha(1+x)+(\beta+2)(1-x)\bigr)P_{\ell-1}^{(\alpha,\beta+2)}(x)
  P_{\ell}^{(\alpha,\beta)}(x)P_{\ell}^{(\alpha-2,\beta+2)}(x)\n
  &=P_{\ell}^{(\alpha,\beta)}(x)P_{\ell}^{(\alpha-2,\beta+2)}(x)\n
  &\qquad\times\Bigl(
  (1+x)\bigl\{2(\ell+\alpha-1)P_{\ell-1}^{(\alpha-1,\beta+3)}(x)
  -2\alpha P_{\ell-1}^{(\alpha,\beta+2)}(x)\bigr\}\n
  &\phantom{\qquad\times\Bigl(}
  +(1-x)\bigl\{2(\ell+\beta+1)P_{\ell-1}^{(\alpha+1,\beta+1)}(x)
  -2(\beta+2)P_{\ell-1}^{(\alpha,\beta+2)}(x)\bigr\}\Bigr)\n
  &=P_{\ell}^{(\alpha,\beta)}(x)P_{\ell}^{(\alpha-2,\beta+2)}(x)
  P_{\ell-2}^{(\alpha+1,\beta+3)}(x)\n
  &\qquad\times(\ell+\alpha+\beta+2)\bigl((1+x)(x-1)+(1-x)(x+1)\bigr)\n
  &=0.
\end{align}
Here we have used Lemma (C),(D) in the curly brackets $\{\cdots\}$.
This concludes the proof of the identity \eqref{jacidenfin1}.

It is well known that the Laguerre polynomial is obtained from the Jacobi
polynomial in the following limit:
\begin{equation}
  \lim_{\beta\to\infty}P_n^{(\alpha,\,\beta)}\bigl(1-\tfrac{2x}{\beta}\bigr)
  =L_n^{(\alpha)}(x).
\end{equation}
When the same limit is applied, the Lemma (C)--(D) reduce to Lemma (A)--(B).
Likewise the cubic identity for the shape invariance of $X_\ell$ Jacobi
polynomial \eqref{jacidenfin1} reduces to that of the $X_\ell$ Laguerre
polynomial \eqref{lagidenfin}, when divided by $-4$ and $\alpha$ is
replaced by $\alpha+1$.

\subsection{hyperbolic DPT}

Here we briefly remark that the shape invariance condition for
the deformed hyperbolic DPT reduces, as expected, to the same identity
as that for the trigonometric DPT \eqref{jacidenfin1} derived above.
We fix $\ell$ and use new parameters $\alpha$ and $\beta$ instead of
$g$ and $h$,
\begin{equation}
  \alpha\eqdef -g-\ell-\tfrac12,\quad \beta\eqdef -h+\ell-\tfrac32.
\end{equation}
By using \eqref{dPn=Pn-1}, the polynomial $\xi_{\ell}$ and its
derivative are expressed just same as \eqref{xil->P}.
The conditions for the shape invariance \eqref{Deltazero} with
\eqref{shapeD2}, after replacing $\eta(x)$ with $x$ and dividing
by $\ell+\alpha+\beta+1$, is simplified to the same
identity \eqref{jacidenfin1} as that for the trigonometric DPT.

\section{Summary and Comments}
\setcounter{equation}{0}

Analytic proofs are provided for the shape invariance  of the recently
derived infinite family of potentials \cite{os16} obtained by deforming
the radial oscillator potential and the trigonometric/hyperbolic
Darboux-P\"oschl-Teller potential \cite{darboux,PT} by a degree
$\ell$ ($\ell=1,2,\ldots$) eigenpolynomial.
The shape invariance conditions are reduced to new polynomial identities
of degree $3\ell$ involving cubic products of the Laguerre
\eqref{lagidenfin} or Jacobi polynomials \eqref{jacidenfin1}.
Then these identities are proved elementarily by combining simple linear
identities \eqref{lemma1}--\eqref{lemma2} among the Laguerre and
\eqref{lemma3}--\eqref{lemma4} among the Jacobi, polynomials of
neighbouring degrees $n,n-1$ and of neighbouring parameters,
$\alpha,\alpha\pm1$ and $\beta,\beta\pm1$.
Even these linear identities seem not widely recognised.

The totality of the eigenvalues and the corresponding eigenfunctions
together with the normalisation constants etc. of these infinite
family of quantum mechanical systems are obtained via the Rodrigues
type formulas \eqref{genformula0} and \eqref{genformula} and reported
as (17)--(21), (29)--(36) and (44)--(46) of \cite{os16}.
In a future publication \cite{hos}, we will present analytic derivation
of various results reported in \cite{os16}.
They include: derivation of equivalent but much simpler looking forms
of the $X_{\ell}$ polynomials together with the normalisation constants,
the verification of the actions of the forward and backward shift
operators on the $X_{\ell}$ polynomials, Gram-Schmidt orthonormalisation
for the algebraic construction of the $X_{\ell}$ polynomials, the analysis
of the second order differential equations for the $X_{\ell}$ polynomials
within the framework of the Fuchsian differential equations in the entire
complex $x$-plane, etc. In these analysis, the linear identities Lemma
(A)--(D) play important r\^oles.
The forward and backward shift relations mentioned above
(eqs.(49),(22),(37),(47) of \cite{os16}) are also reduced to polynomial
identities involving cubic products of the Laguerre or Jacobi polynomials.
We will provide proofs for them in the future publication \cite{hos}, too.

Let us also mention that the same method, deformation in terms of a
degree $\ell$ eigenpolynomial, applied to the discrete quantum mechanical
Hamiltonians for the Wilson and Askey-Wilson polynomials produced
two sets of infinitely many shape invariant systems together with
exceptional ($X_{\ell}$) Wilson and Askey-Wilson polynomials
($\ell=1,2,\ldots$) \cite{os17}.

\section*{Acknowledgements}

This work is supported in part by Grants-in-Aid for Scientific Research
from the Ministry of Education, Culture, Sports, Science and Technology,
No.19540179.


\end{document}